\documentclass[
prl,
twocolumn,
notitlepage
showpacs,
preprintnumbers,
bibnotes,
amsmath,
amssymb,
aps
]{revtex4-1}

\usepackage{enumitem}
\usepackage{graphicx} 
\usepackage{dcolumn} 
\usepackage{bm} 
\usepackage[flushleft]{threeparttable}
\usepackage{braket}
\usepackage{multirow}
\usepackage{array}
\usepackage{url}
\usepackage{booktabs}
\usepackage{float}
\usepackage{hyperref}
\usepackage{mathtools}
\usepackage{setspace}
\usepackage{amssymb,amsfonts,amsmath,amstext}
\hypersetup{colorlinks=true, urlcolor=blue, citecolor=cyan, pdfborder={0 0 0},}
\newcolumntype{P}[1]{>{\centering\arraybackslash}p{#1}}

\usepackage{color,soul}

\newcommand{\nn}{\nonumber \\}
\renewcommand{\v}[1]{{\bm #1}}
\newcommand{\ba}{\begin{eqnarray}}
\newcommand{\ea}{\end{eqnarray}}

\begin{document}
\title{Real-time ab initio simulation of inelastic electron scattering using the exact, density functional, and alternative approaches} 

\author{Yeonghun Lee}
\affiliation{Department of Materials Science and Engineering, University of Texas at Dallas, Richardson, TX 75080, USA}
\author{Xiaolong Yao}
\affiliation{Department of Materials Science and Engineering, University of Texas at Dallas, Richardson, TX 75080, USA}
\author{Massimo V. Fischetti}
\affiliation{Department of Materials Science and Engineering, University of Texas at Dallas, Richardson, TX 75080, USA}
\author{Kyeongjae Cho}
\email{kjcho@utdallas.edu}
\affiliation{Department of Materials Science and Engineering, University of Texas at Dallas, Richardson, TX 75080, USA}
\date{\today} 

\begin{abstract}
To investigate inelastic electron scattering, which is ubiquitous in various fields of study, we carry out ab initio study of the real-time dynamics of a one-dimensional electron wave packet scattered by a hydrogen atom using different methods: the exact solution, the solution provided by time-dependent density functional theory (TDDFT), and the solutions given by alternative approaches. This research not only sheds light on inelastic scattering processes but also verifies the capability of TDDFT in describing inelastic electron scattering. We revisit the adiabatic local-density approximation (ALDA) in describing the excitation of the target during the scattering process along with a self-interaction correction and spin-polarized calculations. Our results reveal that the ALDA severely underestimates the energy transferred in the regime of low incident energy particularly for a spin-singlet system. After demonstrating alternative approaches, we propose a hybrid ab initio method to deal with the kinetic correlation alongside TDDFT. This hybrid method would facilitate first-principles studies of systems in which the correlation of a few electrons among many others is of interest.
\end{abstract}

\maketitle            


\section{Introduction}
\begin{figure}[tb]
\includegraphics{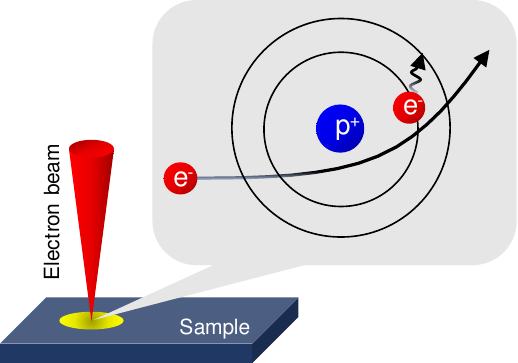}
\centering
\caption{\label{fig:1deh_1}Schematic representation of an inelastic electron scattering process along with an energetic electron impacting a sample.}
\end{figure}

Scattering theory is one of the most fundamental and useful tools in physics. Indeed, inelastic electron scattering by target atoms, molecules, or solids is a demanding problem due to its many-body nature along with the internal degrees of freedom of the excited target. Inelastic electron scattering plays an important role in a wide variety of research fields: electron-beam-induced deposition \cite{engmann_absolute_2013, gonzalez-martinez_electron-beam_2016, sprenger_electron-enhanced_2018}, electron microscopies \cite{meyer_imaging_2008, ciston_surface_2015}, electron radiation damage in semiconductors and metals \cite{corbett_electron_1966, messenger_effects_1992}, hot electron inelastic scattering in devices \cite{ziaja_ionization_2006, tse_ballistic_2008, bernardi_ab_2015}, DNA damaged by electron scattering \cite{boudaiffa_resonant_2000, boudaiffa_cross_2002, caron_low-energy_2003, tonzani_low-energy_2006, garcia_gomez-tejedor_nanoscale_2012, zheng_effective_2018}, electron therapy \cite{kassis_cancer_2003, hogstrom_review_2006}, etc. (see Fig.\,\ref{fig:1deh_1}). Although it is crucial to understand the electronic excitation of the targets during all those inelastic scattering processes, the inherent complexity of the many-body problems hinders a clear interpretation and a proper computation of the dynamics. Moreover, in order for the low-energy scattering measurement to resolve and explain reactivity as well as optical and material properties, it is necessary to develop such computational tools including more representative descriptions of low-energy electron scattering.

Density functional theory (DFT) \cite{hohenberg_inhomogeneous_1964, kohn_self-consistent_1965} has been extensively used in the field of computational physics and chemistry, dealing with many-body problems in an approximate way by replacing the many-electron system with an auxiliary non-interacting Kohn-Sham (KS) system. Subsequently, the time evolution of the KS system can be obtained using time-dependent density functional theory (TDDFT) \cite{runge_density-functional_1984, maitra_perspective:_2016}. The linear-response formalism of TDDFT has been used to obtain the phase shift with respect to an incident electron, focusing on elastic scattering \cite{wasserman_continuum_2005, van_faassen_time-dependent_2007, van_faassen_time-dependent_2009, lacombe_electron_2018}. In contrast to elastic scattering, an inelastic scattering process is nonlinear, and so it is necessary to account for the time-resolved dynamics. In this regard, real-time TDDFT exhibits a better description of the interaction between an energetic electron and matter; moreover, real-time TDDFT has been utilized to simulate the dynamics of inelastic electron scattering events, resulting in the electronic excitation of a target \cite{tsubonoya_time-dependent_2014, miyauchi_electron_2017, ueda_time-dependent_2018, ueda_secondary-electron_2018}.

For a  practical use of TDDFT, the adiabatic local-density approximation (ALDA) \cite{maitra_perspective:_2016} has been adopted reluctantly without sufficient validation as a first-order approximation for the exchange-correlation (XC) functional. However, the ALDA cannot fully capture the nonlinear dynamics of an excited system that does not return to the ground state at each time step. Furthermore, the ALDA is more problematic when dealing with the scattered electron because its energy is far from its ground state. Therefore, efforts have been made to validate the ALDA when describing electron scattering. It has been confirmed that the ALDA XC functional deviates from the exact XC functional \cite{lacombe_electron_2018, suzuki_exact_2017}. This results in the emergence of a nonphysical reflection probability and phase shift of the scattered electron. However, there is no information available about the ability of the ALDA to provide reliable information on the energy transferred during the scattering process. The internal excitation of the target contains critical information, directly connected to subsequent dynamics of the system, such as chemical reactions. Moreover, a systematic validation in a wide range of the incident energy is yet to be obtained.

In this work, we investigate the underlying physics of the energy-transfer process driven by inelastic scattering and, at the same time, to validate the ALDA for the dynamics. To do so, we deal exactly with the real-time dynamics of electron-hydrogen scattering in one dimension (1D e-H scattering) \cite{lappas_computation_1996,lacombe_electron_2018, suzuki_exact_2017}; the exact solution is then compared with the results obtained using the ALDA in TDDFT. The system contains only two electrons and a proton. Although a two-electron system is the extreme limit of a many-electron system, DFT or TDDFT remains effective even for this simplest case of a many-electron system, as illustrated by an example of the helium atom \cite{li_helium_2017, li_comparing_2019}. Furthermore, the exact solution of a simple 1D system, consisting of a very few electrons and atoms, has been exploited to verify nuclear quantum effects \cite{abedi_correlated_2012, suzuki_time-dependent_2018} and the ALDA \cite{lacombe_electron_2018, suzuki_exact_2017, fuks_exploring_2018, elliott_universal_2012} in terms of TDDFT. Hence, the 1D system consisting of an incoming electron and a stationary hydrogen atom would also permit us to validate the ALDA in treating scattering processes. To compare the exact solution with the ALDA in a proper way, spin polarization is considered using the local spin density approximation (LSDA) \cite{oliver_spin-density_1979}. whereas Refs.\,\cite{lacombe_electron_2018, suzuki_exact_2017} pay less attention to the spin polarization. We also discuss the intricacies originating from a fictitious single-electron KS system, such as the time-evolving KS orbital energy level and the final electronic configuration. At the end, we benchmark alternative approaches to simulate electron scattering and propose a hybrid method that may overcome the limitation of the ALDA in treating the inelastic electron scattering. Despite the simplicity of the 1D e-H scattering, the interpretation obtained through this work can be extended to even larger systems with heavier atoms and more electrons while ab initio studies investigate low-energy inelastic scattering phenomena in a wide range of (bio)chemical/physical/materials interests. 

\section{The exact solution}
To study real-time dynamics of 1D e-H scattering, the two-body time-dependent Schrödinger equation (TDSE) $i\partial\Psi(x_1,x_2,t)/\partial t=\hat{H}(x_1,x_2)\mathrm{\Psi}(x_1,x_2,t)$ is solved exactly in a numerical way \cite{lappas_computation_1996,lacombe_electron_2018, suzuki_exact_2017, wachter_numerical_2017}. We use atomic units hereafter unless otherwise stated. The Hamiltonian of a two-electron system is given by
\ba
\hat{H}\left(x_1,x_2\right)=&&-\frac{1}{2}\frac{\partial^2}{\partial x_1^2} -\frac{1}{2}\frac{\partial^2}{\partial x_2^2}+v_{ext}\left(x_1\right) \nn 
&&+v_{ext}\left(x_2\right)+w_{ee}\left(x_1,x_2\right), 
\label{eq:1deh_2}
\ea
where $w_{ee}\left(x_1,x_2\right)=1/\sqrt{\left(x_1-x_2\right)^2+1}$ is the soft-Coulomb interaction and $v_{ext}\left(x\right)=-1/\sqrt{\left(x-x_H\right)^2+1}$ is the soft-Coulomb potential induced by the stationary hydrogen nucleus at $x_H$. In the soft-Coulomb potential, cusps disappear with removing the singularity at zero separation. Ref.\,\cite{baker_one-dimensional_2015} has constructed an exponential interaction, which takes into account cusp of the potential at the origin. Although compared with the soft-Coulomb form, the exponential interaction greatly reduces computational cost for calculating accurate quantities with the density matrix renormalization group, they exhibit a similar quality in terms of energy components for various systems. The softness parameter in the denominator can be determined according to the particular application of study. The soft-Coulomb interaction with the softness parameter of unity has been verified in terms of the local-density approximation (LDA), where 1D molecules with this softness parameter qualitatively mimic three-dimensional (3D) molecules well \cite{helbig_density_2011}. Hence, the 1D soft-Coulomb interaction adopted here is expected to capture the essential physics of 1D e-H scattering.

Having determined the initial wavefunction, we let it evolve according to the TDSE in order to study the dynamics. The initial wavefunction $\mathrm{\Psi}\left(x_1,x_2,t=0\right)$ can be chosen as the spin-singlet or spin-triplet state given by the Slater determinant of the hydrogen ground state $\psi_1(x)$ and a Gaussian wave packet $\psi_{WP}\left(x\right)$, expressed as 
\ba
\mathrm{\Psi}\left(x_1,x_2,t_0\right)=\frac{1}{\sqrt2}&&\left[\psi_1\left(x_1\right)\psi_{WP}\left(x_2\right)\right. \nn 
&&\left.\pm\psi_{WP}\left(x_1\right)\psi_1\left(x_2\right)\right],
\ea
where the $+$ and $-$ signs correspond to singlet and triplet states, respectively. The Gaussian wave packet $\psi_{WP}\left(x\right)$ is given by 
\ba
\psi_{WP}\left(x\right)=&&\left(\frac{1}{\pi\sigma^2}\right)^{1/4} \nn 
&&\times \exp{\left[-\frac{\left(x-x_0\right)^2}{2\sigma^2} 
+ip(x-x_0)\right]},
\ea
with momentum p, position $x_0$, and width $\sigma$. Throughout this paper, we use $x_H=0$ a.u., $x_0=-20$ a.u., and $\sigma=2.236$ a.u.\,that is equivalent to $\alpha=0.1$ in Refs.\,\cite{lacombe_electron_2018, suzuki_exact_2017} (see Fig.\,S1 in Supplemental Material \cite{suppl} for results with different values of $\sigma$).

\begin{figure}[tb]
\includegraphics{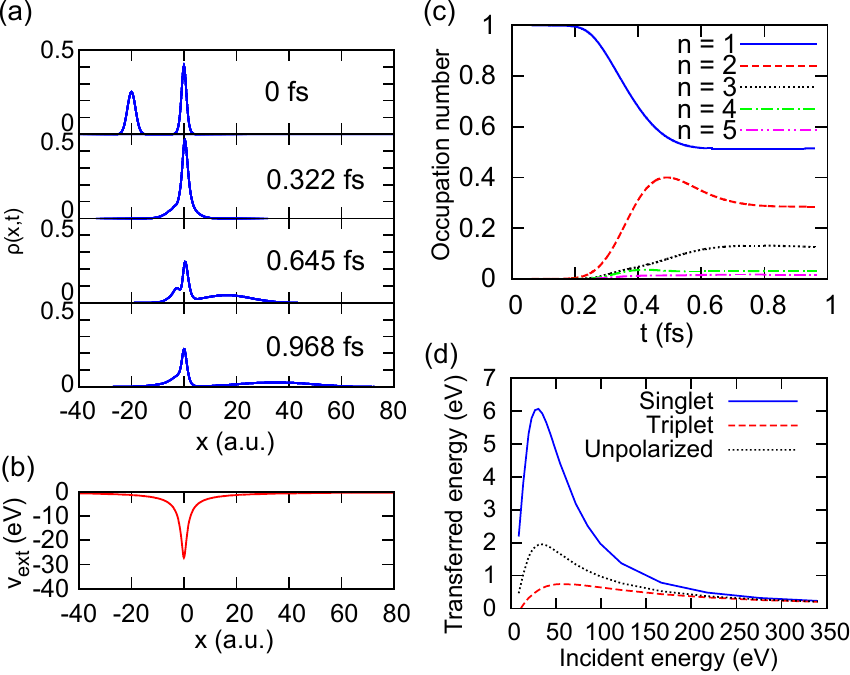}
\centering
\caption{\label{fig:1deh_2}Real-time dynamics as obtained from the exact solution. (a) Time evolution of electron density ($t=0.322,0.645,0.968$ fs). At the initial time, peaks on the left- and right-hand sides indicate the incoming electron wave packet and the electron bounded by the proton, respectively. (b) External potential induced by the proton, assumed to be stationary at the origin. (c) Time evolution of the occupation number for each hydrogen orbital n. In (a–c), the initial momentum of the incident electron is 1.5 a.u., which corresponds to a kinetic energy of 30.6 eV, and the spin state of the initial wavefunction is set to the singlet state. (d) Transferred energy as a function of the incident energy when the initial spin state is the singlet (solid line), triplet (dashed line), or unpolarized state (dotted line).}
\end{figure}

Figure\,\ref{fig:1deh_2}(a) presents the time-evolving electron density of the incomming wave pacekt and the bound electron under the external potential shown in Fig.\,\ref{fig:1deh_2}(b) (the corresponding animation is given in Supplemental Material \cite{suppl}). Given a two-electron wavefunction, the electron density is calculated by using
\ba
\rho\left(x,t\right)=\int{dx^\prime\left|\mathrm{\Psi}\left(x,x^\prime,t\right)\right|^2}+\int{dx^\prime\left|\mathrm{\Psi}\left(x^\prime,x,t\right)\right|^2}.
\ea
As time evolves, the wave packet broadens and splits into transmitted and reflected waves, and the bound-electron density also broadens and exhibits oscillations. The broadening of the bound-electron density is evidence of the energy transfer from the incident electron to the bound electron. This is caused by an electron transition from the ground state to the first or higher excited state. Figure\,\ref{fig:1deh_2}(c) illustrates the time-evolving occupation numbers during the inelastic scattering process. The occupation number of the n-th energy level can be calculated by using
\ba
\left|c_n\left(t\right)\right|^2=&&\int{dx_1\left|\int{dx_2\psi_n^\ast\left(x_2\right)\mathrm{\Psi}\left(x_1,x_2,t\right)}\right|^2} \nn 
&&+\int{dx_2\left|\int{dx_1\psi_n^\ast\left(x_1\right)\mathrm{\Psi}\left(x_1,x_2,t\right)}\right|^2}.
\ea
where $\psi_n$ is the $n$-th hydrogen orbital. Obviously, such a definition is not rigorous, since the incident electron and the bound electron are indistinguishable and $\mathrm{\Psi}\left(x_1,x_2,t\right)$ contains both. Nonetheless, we can approximately distinguish them in energy space. Appropriately choosing the upper limit of $n$ enables us to extract information about the bound electrons from the two-electron wavefunction. Combining Figs.\,\ref{fig:1deh_2}(a, c), it can be observed that internal excitation of the target hydrogen atom occurs when the two-electron densities spatially overlap (this occurs for times approximately between 0.2 and 0.7 fs). An interesting event is observed from the real-time dynamics. The second excited state ($n=3$) becomes occupied after the first excited state ($n=2$) is populated. This stems from the fact that the transition from the first to second excited state is a primary process for populating the second excited state.

An important phenomenon seen during inelastic scattering is the transfer of energy. Taking into account the electronic excitation of the hydrogen atom, the amount of energy transferred is given by an average weighted by the occupation numbers:
\ba
E_{trans}=\sum_{n=2}{\left|c_n\left(t\right)\right|^2\left(\varepsilon_n-\varepsilon_1\right)},
\ea
where $\varepsilon_n$ is the $n$-th energy level of the target hydrogen atom. We perform spin-singlet, spin-triplet, and spin-unpolarized calculations. The spin-unpolarized results are obtained by combining the singlet and triplet results in a ratio of 1:3. It is known that the 2s state of parahelium (singlet) exhibits a higher energy level than that of orthohelium (triplet) because the spatially symmetric wavefunction of the singlet state gives rise to strong Coulomb repulsion. In the same manner, the energy transfer for the singlet state is more significant than that of the triplet state due to the stronger Coulomb interaction [Fig.\,\ref{fig:1deh_2}(d)]. It is also shown that peaks of the transferred energy appear around the incident energy of 35 eV. These transferred-energy peaks can be explained as follows: An incident energy that is too low cannot excite the target in any significant way; on the contrary, at a very high incident energy interacting time decreases, because of the high speed of the incoming electron. This dependence on the incident energy can also be found by looking at the dependence of the scattering cross-section on the initial and final electron wavenumbers, $k$ and $k^\prime$, as seen in the pre-factor on the right-hand side of Eq.\,(\ref{eq:1deh_23}) in the section IV. When exciting molecules to accelerate chemical reactions, it seems problematic that only a small portion of incident energy is transferred to a target molecule regardless of how high the energy of the incident electron is. For example, the maximum transferred-energy seen in 1D e-H scattering when performing spin-unpolarized calculations reaches only 2 eV [Fig.\,\ref{fig:1deh_2}(d)], which is smaller than the energy gap between the ground and the first excited states (see Fig.\,\ref{fig:1deh_3}(a) for the energy gap). Considering the transferred energy we calculate is a probabilistic expectation value of all possible outcomes, any excitation can occur with some probability corresponding to the occupation number displayed in Fig.\,\ref{fig:1deh_2}(c). Although the probability is small, the scattering process provides excitation high enough to facilitate subsequent chemical reactions.

\section{Time-dependent density functional theory}

\begin{figure}[tb]
\centering
\includegraphics{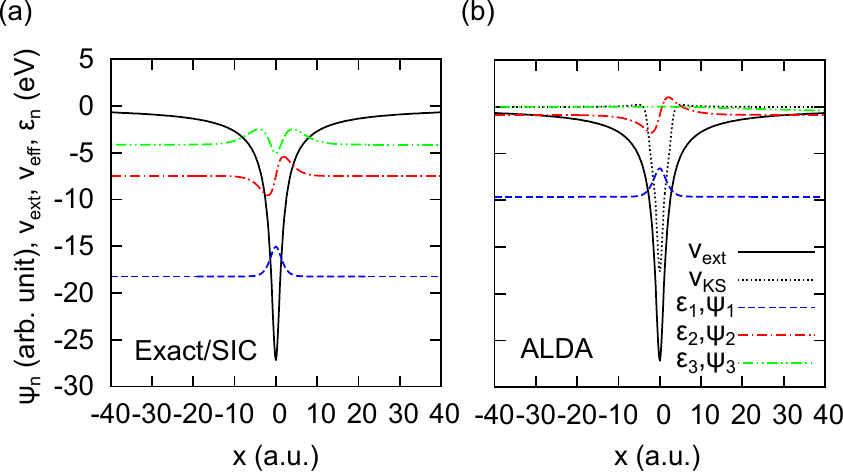}
\caption{\label{fig:1deh_3}The lowest three energy levels and orbitals in (a) the exact 1D hydrogen and (b) the corresponding KS system. The solid line represents the external potential. The dotted line in (b) indicates the effective KS potential $v_{KS}=v_{ext}+v_{Hartree}+v_{xc}$. Since deviation in the KS system originates from self-interaction, the SIC corrects KS orbitals to the exact ones.}
\end{figure}

Based on TDDFT, we solve the time-dependent KS equation $i\partial\phi\left(x,t\right)/\partial t={\hat{H}}_{KS}\phi\left(x,t\right)$ for each KS orbital, describing the dynamics of the incident electron and the bound hydrogen electron. The KS Hamiltonian is given by
\ba
{\hat{H}}_{KS}=-\frac{1}{2}\frac{\partial^2}{\partial x^2}+v_{ext}\left(x\right)+v_{Hartree}\left[\rho\right]+v_{xc}\left[\rho\right],
\ea
where $v_{Hartree}\left[\rho\right]$ is the Hartree potential, $v_{xc}\left[\rho\right]=\delta E_{xc}\left[\rho\right]/\delta\rho\left(x,t\right)$ is the XC potential, and $\rho\left(x,t\right)=\left|\phi_H\left(x,t\right)\right|^2+\left|\phi_{WP}\left(x,t\right)\right|^2$ is the electron density. The ALDA XC potential is expressed as 
\ba
v_{xc}\left[\rho\left(x^\prime,t^\prime\right)\right]\left(x,t\right)&&=v_{xc}^{LDA}\left(\rho(x,t)\right) \nn 
&&=\varepsilon_{xc}\left(\rho(x,t)\right)+\frac{\partial\varepsilon_{xc}\left(\rho\left(x,t\right)\right)}{\partial\rho\left(x,t\right)},
\ea
where $\varepsilon_{xc}$ is a 1D XC energy density \cite{helbig_density_2011, wagner_reference_2012}. Note that the ALDA washes out the memory at $t^\prime<t$. The initial KS states are set to $\phi_H\left(x,t_0\right)=\psi_1\left(x\right)$ and $\phi_{WP}\left(x,t_0\right)=\psi_{WP}\left(x\right)$. When trying to compare the results of the exact and TDDFT approaches, in principle we face the issue of determining the initial state of the hydrogen electron, since the KS orbital is not the exact ground state (Fig.\,\ref{fig:1deh_3}). However, we have verified that the use of two different initial ground states—the exact 1D hydrogen orbital and the corresponding KS orbital—has resulted in a very similar inelastic scattering behavior, thanks to the fact that  the electron density given by  the exact orbital is almost identical to the density associated with the KS orbital (see Fig.\,S2 in Supplemental Material \cite{suppl}).

\begin{figure}[tb]
\centering
\includegraphics{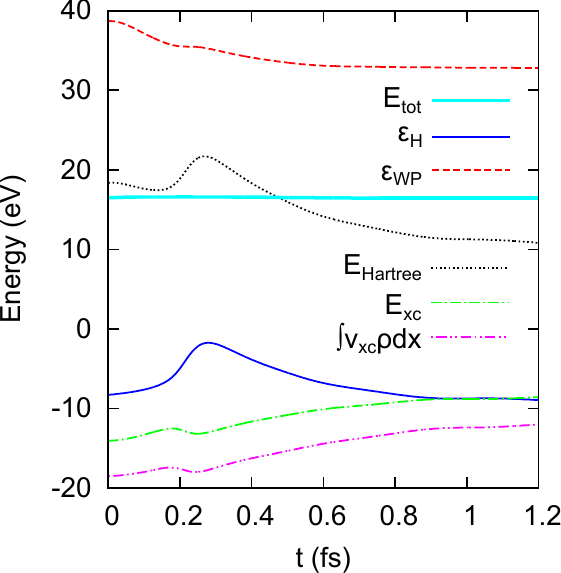}
\caption{\label{fig:1deh_4}Time evolution of energy terms pertaining to the KS system: total energy $E_{tot}$ (thicker solid line), KS orbital energies for the bound hydrogen electron $\varepsilon_H$ (thinner solid line) and electron wave packet $\varepsilon_{WP}$ (dashed line), Hartree energy $E_{Hartree}$ (dotted line), exchange-correlation energy $E_{xc}$ (dash-single dotted line), and exchange-correlation potential integral (dash-double dotted line). Note that the KS orbital energies are expectation values of the KS Hamiltonian for those time-evolving KS orbitals. Spin-unpolarized calculation is adopted all the way through.}
\end{figure}

The conservation of energy implied by a time-invariant Hamiltonian is a fundamental concept of physics. Despite the fact that the  KS Hamiltonian of each electron depends on a time-evolving density, the total energy of the KS system is conserved when the XC functional is adiabatic and local as in the ALDA \cite{schleife_plane-wave_2012}. The total energy of a KS system for 1D e-H scattering is given by
\ba
E_{tot}=&&\varepsilon_H+\varepsilon_{WP}+E_{Hartree}\left[\rho\right]+E_{xc}\left[\rho\right] \nn 
&&-\int{v_{xc}\left[\rho\right]\rho\left(x,t\right)dx}.
\ea
The fact that the total energy shown in Fig.\,\ref{fig:1deh_4} is constant additionally shows that the real-time integration for the time-dependent KS system is numerically stable and accurate. In addition, Fig.\,\ref{fig:1deh_4} shows the time evolution of the KS orbital energies, $\varepsilon_H$ and $\varepsilon_{WP}$, which are expectation values of the KS Hamiltonian on the states corresponding to the hydrogen atom and to the incident wave packet, respectively: $\varepsilon_H=\left\langle\phi_H\middle|{\hat{H}}_{KS}\middle|\phi_H\right\rangle$ and $\varepsilon_{WP}=\left\langle\phi_{WP}\middle|{\hat{H}}_{KS}\middle|\phi_{WP}\right\rangle$. It is known that the KS orbital energy level does not have any physical meaning; therefore, the energy change of the KS orbital over time is not equal to the energy change of each system, even when the incident electron and the hydrogen are sufficiently separated in space, as it happens at late times. Hence, the final $\varepsilon_H$ can be lower than the initial $\varepsilon_H$, although obviously, the hydrogen atom gains some energy from the incident electron. 

\begin{figure}[tb]
\centering
\includegraphics{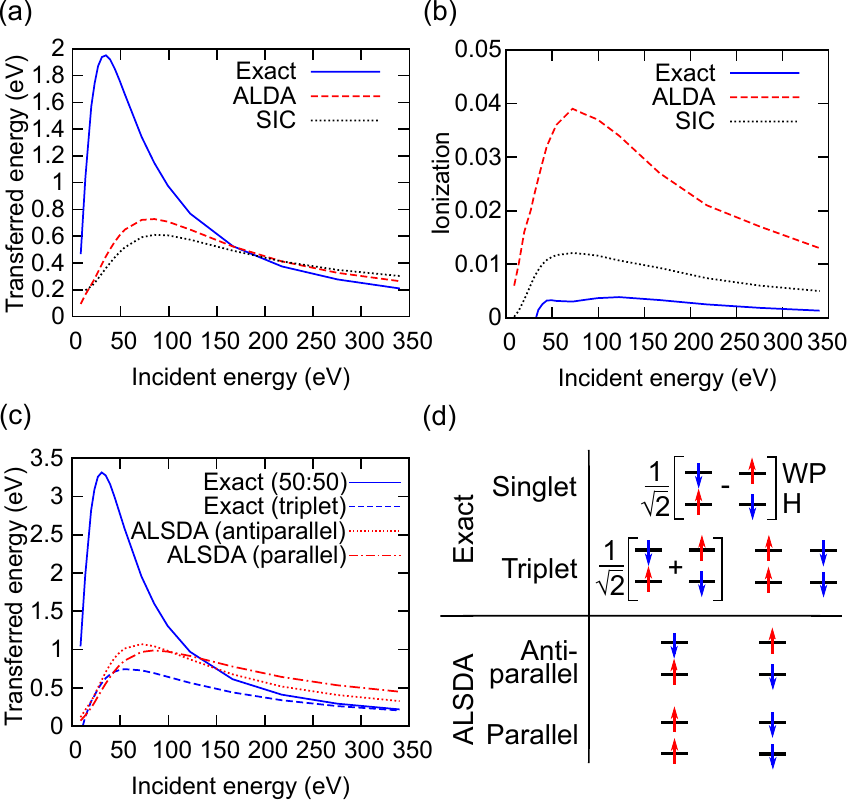}
\caption{\label{fig:1deh_5}TDDFT results of the spin-unpolarized and spin-polarized calculations. (a) Transferred energy and (b) ionization probability calculated using the exact solution (solid line), TDDFT with the ALDA (dashed line), and TDDFT with SIC (dotted line). The results shown in (a, b) have been obtained using spin-unpolarized calculations. (c) Transferred energy for different spin states shown in (d). An equal mixture (50:50) of the singlet and triplet states in the exact solution corresponds to antiparallel spin states in the ALSDA \cite{ziegler_calculation_1977, cramer_density_1995, ess_singlettriplet_2011}. (d) Possible spin states of a two-electron system. The upper and lower states are for the incident wave packet electron (WP) and the bound electron in the hydrogen atom (H), respectively.}
\end{figure}

Figures\,\ref{fig:1deh_5}(a,b) show that the ALDA largely underestimates the transferred energy, whereas the ionization probability is overestimated by the ALDA. The ionization probability is evaluated by counting the number of bound electrons that occupy the orbitals below the vacuum level. The deviations are caused by the fact that the ALDA does not properly reproduce correlation, which plays a crucial role especially in describing the low-energy scattering \cite{lappas_computation_1996,lacombe_electron_2018, suzuki_exact_2017, fuks_exploring_2018}. In most XC functionals for the DFT, the self-interaction error originating from the spurious interaction of an electron with its own mean-field is one of the major sources of error. The self-interaction error can be cured by implementing the self-interaction correction (SIC), where the contribution of an electron interacting with itself is subtracted from the $v_{Hartree}\left[\rho\right]$ and $v_{xc}\left[\rho\right]$ \cite{perdew_self-interaction_1981}. It turns out that the overestimation of ionization probability largely originates from a fewer number of bound states induced by the self-interaction error (see Fig.\,\ref{fig:1deh_3} for shallower bound states induced by the self-interaction error for the ALDA).  As we expected, the result from the SIC is in a better agreement with the exact solution than the ALDA, concerning the ionization probability [Fig.\,\ref{fig:1deh_5}(b)]. The SIC also repairs the poor delocalization of the wave packet caused by the self-interaction error (see Fig.\,S3 in Supplemental Material \cite{suppl}). However, the underestimation of transferred energy is still not corrected by the SIC. The SIC does not properly capture kinetic correlation, which is missing in the ALDA.

Whereas so far we have discussed only spin-unpolarized calculations, now we employ the adiabatic local spin density approximation (ALSDA) to account for a spin-polarized system. Figure\,\ref{fig:1deh_5}(d) shows a schematic diagram of the different spin states of a two-electron system, in which the two electrons occupy different energy levels. The parallel spins in the ALSDA correspond to the triplet state in the exact solution, and the antiparallel spins in the ALSDA correspond to an equal mixture (50:50) of the singlet and the triplet states in the exact solution \cite{ziegler_calculation_1977, cramer_density_1995, ess_singlettriplet_2011}. The equal mixture can also be regarded as a virtual system of distinguishable electrons. The antiparallel-spin corresponding to the artificial system brings about so-called spin contamination, in which the Slater determinant formed by the KS wavefunctions is not an eigenstate with respect to the square of the total spin angular momentum $\hat{S}^2$ \cite{ess_singlettriplet_2011}. This is an inherent error of the conventional DFT. Even when ignoring the spin contamination, the antiparallel-spin state is adopted to compare the ALSDA with the exact solution. Figure\,\ref{fig:1deh_5}(c) shows that while the exact solution differentiates clearly the antiparallel- and the parallel-spin states, the difference almost disappears in the ALSDA using the equal mixture of the singlet and triplet states. This means that the adiabatic XC functional cannot capture correctly the strong interaction present in the singlet state but can capture relatively well the moderate interaction that is present in the triplet state. 

Regarding the problem of estimating the energy transferred in the collision, on the one hand, in analyzing the exact solution, it is straightforward to extract transferred energy between the incident electron and the hydrogen atom by simply projecting the initial and the final wavefunctions onto the exact hydrogen orbitals. Moreover, the projection readily provides information about the electronic occupation of the final state induced by the inelastic scattering. On the other hand, when using the KS approximation, we use a different way to calculate the transferred energy, since an absolute value of the KS orbital-energy does not correspond directly to the energy of the system. When the interaction between the two electrons is negligible, we can describe them separately in the KS Hamiltonian by excluding one electron. This technique is valid when considering the initial state and the final state, in which the bound and the incident electrons are separated spatially; therefore, we can just remove the incident wave packet from the system to calculate the energy of the KS hydrogen system, $E_H\left(t\right)$. Note that the energy of the KS system, $E_H\left(t\right)$ is not identical to the KS orbital energy level, $\varepsilon_H\left(t\right)$. Hence, transferred energy is defined as the difference between the initial and final energies, 
\ba
E_{trans}=\Delta E_H=E_H\left(t_f\right)-E_H\left(t_0\right).
\label{eq:1deh_Etrans}
\ea
The results obtained using Eq.\,(\ref{eq:1deh_Etrans}) are shown in Fig.\,\ref{fig:1deh_5}. 

\begin{figure}[tb]
\centering
\includegraphics{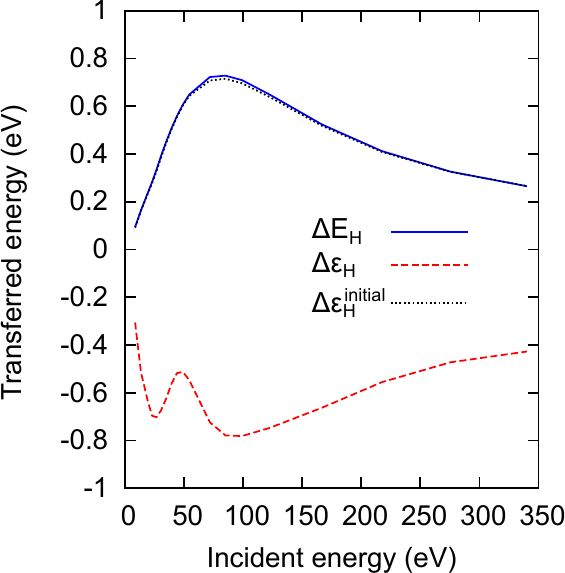}
\caption{\label{fig:1deh_6}Transferred energy calculated using different approaches for the ALDA. The solid line corresponds to the change of the total energy of the hydrogen, $\Delta E_H$. The dashed line shows the difference between the initial and final KS orbital energies of the hydrogen electron, $\Delta \varepsilon_H$. To obtain the final electronic configuration, we can also use the initial ground-state KS orbitals for projection, and the dotted line represents the energy change of the hydrogen in terms of the initial ground-state KS orbitals, $\Delta \varepsilon_H^{initial}$.}
\end{figure}

The final electronic configuration is of interest because it is crucial for the subsequent dynamics driven by the scattering process. There could be two available basis sets expressing electronic structure: KS eigenstates of the final excited state or KS eigenstates of the initial ground state. According to the basis set, one can extract the energy change of the target:
\ba
\Delta \varepsilon_H=&&\varepsilon_H\left(t_f\right)-\varepsilon_H\left(t_0\right) \nn 
=&&\sum_{n}{\left|\left\langle\psi_n\left(t_f\right)\middle|\phi_H\left(t_f\right)\right\rangle\right|^2\varepsilon_n\left(t_f\right)} \nn 
&&-\sum_{n}{\left|\left\langle\psi_n\left(t_0\right)\middle|\phi_H\left(t_0\right)\right\rangle\right|^2\varepsilon_n\left(t_0\right)}, \\
\Delta \varepsilon_H^{initial}=&&\sum_{n=2}{\left|\left\langle\psi_n\left(t_0\right)\middle|\phi_H\left(t_f\right)\right\rangle\right|^2\left[\varepsilon_n\left(t_0\right)-\varepsilon_1\left(t_0\right)\right]}, \nn 
\ea
where $\Delta \varepsilon_H$ is the difference between the initial and final KS orbital energies of the hydrogen electron, and $\Delta \varepsilon_H^{initial}$ is the energy change of the hydrogen, projecting onto the initial ground-state KS orbitals. The eigenvalues $\varepsilon_n\left(t\right)$ and hydrogen orbital basis $\psi_n\left(t\right)$ are obtained by diagonalizing KS Hamiltonian at a given time. Although the change of KS orbital energies at different moments, $\Delta \varepsilon_H$, hardly gives useful information (dashed line in Fig.\,\ref{fig:1deh_6}), the relative level of the KS orbital energy at a given time can still survive as useful information. A similar argument has been made for bandgap or ionization potential \cite{stowasser_what_1999}. In this sense, the time-evolving wavefunctions even at different times can be projected onto the initial ground-state KS orbital basis to keep track of change in the electronic configuration. Figure\,\ref{fig:1deh_6} shows that the energy change of the hydrogen atom given by the projection onto the initial KS orbitals, $\Delta \varepsilon_H^{initial}$, is in good agreement with the transferred energy $\Delta E_H$. This is due to the fact that the basis set of the initial KS orbitals describes well the final electronic configuration along with excitation. Even if the ions were free to move, the ionic position would hardly change during the electron scattering process that lasts only a femtosecond or so. Therefore, the projection onto the initial ground-state KS orbital basis is not only simpler than using the basis at later times but also able to capture the final electronic configuration.

\section{Alternative approaches}
The exact solution is practically intractable for a realistic system and, as aforementioned, the ALDA underestimates substantially the energy transferred at low incident energy. Other approximations often fail to reproduce the interaction of a low-energy electron with a target. Nevertheless, low-energy scattering processes are of interest in many cases. For instance, electrons at low incident energies ($<50$ eV) are responsible for electron-induced DNA damage \cite{boudaiffa_resonant_2000, boudaiffa_cross_2002, caron_low-energy_2003, tonzani_low-energy_2006, garcia_gomez-tejedor_nanoscale_2012, zheng_effective_2018}, and an electron-enhanced atomic layer deposition technique utilizes low incident energies (25\textendash 200 eV) to stimulate surface reactions \cite{engmann_absolute_2013, sprenger_electron-enhanced_2018}.

One may wonder whether or not there is any other way to estimate the effects of low-energy electron-scattering in a computationally feasible way. The time-dependent Hartree-Fock approach \cite{stich_tdhf_1985} cannot be a solution because of the kinetic correlation missing in the ALDA \cite{lacombe_electron_2018, suzuki_exact_2017, fuks_exploring_2018}. This kinetic correlation cannot be captured by the Hartree-Fock method either. The R-matrix theory is a convenient and efficient tool where a system is divided into an internal region and an external region \cite{burke_electron_1971}. In the internal region, short-range interactions are considered as confined; in the external region, the scatteringwave function is approximated by its asymptotic form. The R-matrix theory can describe elastic and inelastic electron scattering from complex large molecules, such as DNA and RNA, where the theory can be combined with ab initio methods within static exchange and neglect of correlation \cite{tonzani_low-energy_2006}. The TDDFT approach enables the incorporation of time-dependent exchange and correlation effects as a density functional in a cost-effective way \cite{van_faassen_time-dependent_2007, krueger_autoionizing_2009}. Furthermore, the TDDFT can provide real-time dynamics, which is not accessible in the R-matrix theory. One can think of time-dependent perturbation theory or consider the classical limit of an incident electron, keeping a low computational cost. Also, we would be able to use the relationship between the exact and the ALDA results shown in the previous section. We investigate whether these methods are feasible in describing the inelastic electron scattering. 

\begin{figure}[tb]
\centering
\includegraphics{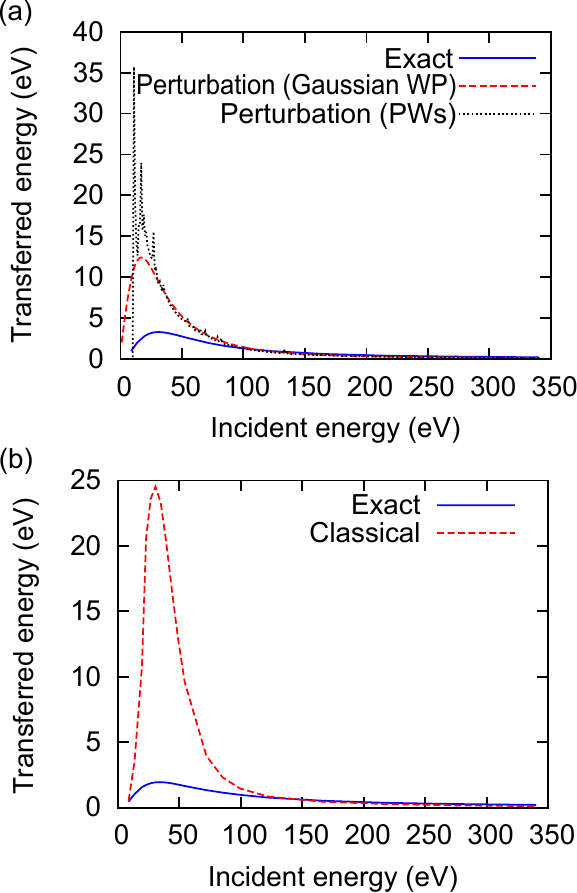}
\caption{\label{fig:1deh_7}Alternative methods to simulate transferred energy driven by the inelastic electron scattering. (a) Transferred energy calculated by using perturbation theory (dashed line), perturbation theory but with plane waves (PWs) (dotted line), and the exact solution (solid line). In the exact solution here, the initial two electrons are assumed to be distinguishable as the same assumption is introduced in the solution derived with the perturbation theory. (b) Transferred energy calculated by taking the classical particle limit with respect to the incoming electron (dashed line), compared with the spin-unpolarized exact solution (solid line).}
\end{figure}

Fermi’s golden rule, which is equivalent to the first-order Born approximation, can be used to calculate the cross-section of inelastic electron-atom scattering \cite{sakurai_modern_1994}. Fermi’s golden rule is derived from the time-dependent perturbation theory. Assuming that the incident electron is distinguishable from the bound electrons, as in Ref.\,\cite{sakurai_modern_1994}, and introducing the soft-Coulomb potential, we can derive the 1D inelastic scattering cross-sections associated with an incident plane wave $e^{ikx}$ and the $n$-th excitation of a target as:
\ba
\sigma_{0\rightarrow n}\left(k\right)=\frac{Z^2}{kk^\prime}&&\left\{\left|\left[-\delta_{n0}+F_n\left(q_+\right)\right]\int{\frac{e^{iq_+x}}{\sqrt{x^2+1}}dx}\right|^2\right. \nn 
&&\left.+\left|\left[-\delta_{n0}+F_n\left(q_-\right)\right]\int{\frac{e^{iq_-x}}{\sqrt{x^2+1}}dx}\right|^2\right\}, \nn 
\label{eq:1deh_23}
\ea
where $Z$ is the atomic number of the target atom, $k^\prime=\sqrt{k^2-2\left(E_n-E_0\right)}$ is the final-state wave-vector of the incident plane wave, and $q_\pm=k\mp k^\prime$ is the difference of wave vector between the incoming and outgoing waves. The first and second terms on the right-hand side represent the forward and backward scattered waves, respectively. Here, $F_n\left(q\right)$ is the form factor, expressed as $ZF_n\left(q\right)=\left\langle n\middle|\sum_{i} e^{iqx_i}\middle|0\right\rangle$, where $i$ is the index of the electrons in the target. The form factor reduces to $F_n\left(q\right)=\left\langle n\middle| e^{iqx}\middle|0\right\rangle$ when dealing with the hydrogen atom. In the e-H scattering, the target energy $E_n$ and the target state $\left.|n\right\rangle$ for the $n$-th excited state can be simplified to $\varepsilon_{n+1}$ and $\psi_{n+1}$ for the $(n+1)$-th energy level of the hydrogen atom. It seems impossible to define a cross-sections in a 1D system, in which every incident particle encounters the scattering center. However, considering the analogy of the relationship $1/\tau_{0\rightarrow n}\left(\v{k}\right)=\sigma\left|\v{j}\right|$ in a 3D system [$1/\tau_{0\rightarrow n}\left(\v{k}\right)$ is a scattering rate, $\sigma$ is a cross-section, and $\v{j}$ is an electron flux], the 1D cross-section simply defines a scattering probability rather than a cross-section \textit{per se} \cite{barlette_integral_2001}. Finally, we can write the transferred energy as 
\ba
E_{trans}=\sum_{n}\int{{\widetilde{\psi}}_{WP}^\ast\left(k\right)\left(E_n-E_0\right)\sigma_{0\rightarrow n}\left(k\right){\widetilde{\psi}}_{WP}\left(k\right)dk}, \nn 
\ea
where ${\widetilde{\psi}}_{WP}\left(k\right)$ is the Fourier transform of a Gaussian wave packet in position space, $\psi_{WP}\left(x\right)$. 

Figure\,\ref{fig:1deh_7}(a) shows the transferred energy using time-dependent perturbation theory. This overestimates severely the energy transferred by an electron with an incident energy lower than 100 eV. The perturbation theory assumes a small change of the incoming wave after scattering. This assumption is fulfilled when incident electron energy is large compared to the strength of scattering potential. For inelastic scattering channels, the incoming wave can be modified substantially by losing a large portion of its kinetic energy, and the assumption fails. As a result, for the inelastic scattering channels, perturbation theory fails to provide a correct scattering probability for high-energy excitations, overestimating their scattering probability. This leads to the overestimation of the energy transferred in a low incident energy. 

Here, we suppose the incident electron is a classical particle and apply Ehrenfest dynamics \cite{todorov_time-dependent_2001, parandekar_detailed_2006} for the incident electron. The Ehrenfest dynamics is obtained by applying the Ehrenfest theorem to a highly localized state in space. We have confirmed the conservation of the total energy, consisting of the classical incident electron and the hydrogen atom, while running the real-time simulation. Figure\,\ref{fig:1deh_7}(b) shows that the classical incident electron gives rise to great overestimation of energy transferred in incident energy lower than 100 eV. This deviation has to do with the extremely localized classical electron creating substantially strong time-dependent field than an electron in a wave packet form. One may expect that this result could also be reproduced by using a wave packet with a small enough $\sigma$, which makes the initial wave packet narrow. However, it is not the case because the width of a free wave packet spreads as $\sqrt{\sigma^2+t^2/\sigma^2}$ \cite{hashimoto_out--time-order_2017}, i.e., an originally narrow wave packet is rapidly delocalized.

\begin{figure}[tb]
\centering
\includegraphics{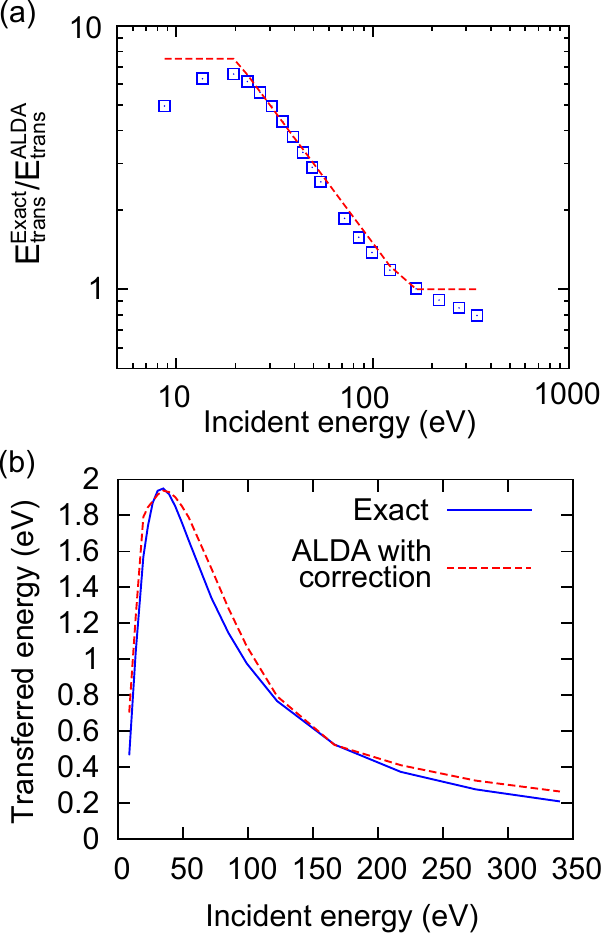}
\caption{\label{fig:1deh_8}(a) The ratio of the transferred energy in the exact solution, $E_{trans}^{ALDA}$, to the transferred energy in the ALDA, $E_{trans}^{ALDA}$ (open squares). In this log-log plot, the dashed line is a function expressed as: $y=150/20$ for $x<20$; $y=150/x$ for $20\le x<150$; $y=150/150$ for $150\le x$. (b) The relationship displayed in (a) is used to correct the transferred energy underestimated in the ALDA. The maximum transferred energy and the incident energy with the maximum point are consistent with the exact values. Spin-unpolarized calculation is adopted.}
\end{figure}

Taking a ratio $E_{trans}^{Exact}/E_{trans}^{ALDA}$, we measure a degree of kinetic correlation as a function of incident energy. Figure\,\ref{fig:1deh_8}(a) shows that kinetic correlation plays a crucial role in the deviation of the energy transferred with the ALDA when the kinetic energy of an incident electron is smaller than 150 eV. Here, the ratio is inversely proportional to the incident energy. This inverse proportion is related to overlapping time of two electrons, since the kinetic correlation is significant when the incoming electron overlaps with the bound electron. In the regime of incident energy higher than 150 eV, the correlation time would be too short to give an emergence of the kinetic correlation. At the turning point at the incident energy of 20 eV, the ratio saturates and starts to decrease as incident energy decreases. This is because inelastic scattering channels open only if the incoming electron energy is larger than the corresponding excited energy of a target. However, the incident energy at the turning point is larger than the first excited energy of the hydrogen atom, where the excitation energy required is 10.7 eV in the exact solution and 8.8 eV in the LDA. Considering a Gaussian incoming wave packet adopted here, the turning point can be shifted to an energy larger than the first excited energy of a target because the Gaussian wave packet is decomposed into plane waves with a certain range of kinetic energy around the given incident energy. 

The expression of the dashed line shown in Fig.\,\ref{fig:1deh_8}(a) can be exploited to correct the transferred energy calculated using the ALDA [Fig.\,\ref{fig:1deh_8}(b)]. Once determining the upper and lower turning points of incident energy, one can make a correction to the scattering probability given by the ALDA. The lower turning point is associated with the first excited energy, and the upper turning point would be associated with the size of target. Although this correction is not so accurate, it can be of practical use to have the quantitatively meaningful scattering probability from the ALDA (see Fig.\,S4 in Supplemental Material \cite{suppl} for effects of the deviation of the parameters on the correction).

In closing, we propose a hybrid TDDFT-TDSE method to resolve the severely underestimated scattering probability, which is a fundamental limitation imposed by the ALDA. Once we accept that DFT or TDDFT well captures the electronic structure of the target, the remaining problem becomes the way of introducing the effects of the incident electron. In a combined Hamiltonian, we separately deal with the two interactions: (1) interaction among bound electrons by using TDDFT; (2) interaction between the incident electron and one of the bound electrons by using TDSE. The hybrid method might be able to reduce computational cost without loss of correlation between the incident electron and the target electrons. The Hamiltonian of this hybrid system is introduced in
\ba
{\hat{H}}_{hybrid}=\sum_{i}^{target}\biggl[-\frac{1}{2}\frac{\partial^2}{\partial x_i^2}+v_{ext}\left(x_i\right)+v_{Hartree}\left[\rho^\prime\right] \nn 
+v_{xc}\left[\rho^\prime\right]+w_{ee}\left(x_i,x_j\right)\biggr]-\frac{1}{2}\frac{\partial^2}{\partial x_j^2}+v_{ext}\left(x_j\right), \nn 
\label{eq:1deh_25}
\ea
where $\rho^\prime$ is target electron density without the incident electron wave packet, given by $\rho^\prime\left(x,t\right)=\sum_{i}^{target}\left|\phi_i(x,t)\right|^2$. For a special case where electrons are assumed to be distinguishable, $j$ is an index only for the incident electron, whereas the incident electron index $j$ is interchangeable with a target electron index $i$ for a general case where the incident electron is indistinguishable from the target electrons. When we take into account e-H scattering along with SIC, Eq.\,(\ref{eq:1deh_25}) is readily reduced to the exact Hamiltonian shown in Eq.\,(\ref{eq:1deh_2}). In doing so, the hybrid TDDFT-TDSE method would facilitate reliable modeling of electron scattering by even more complex targets in terms of the first-principles calculation. As well as electron scattering, this approach can be applied to different systems where we are especially interested in the correlation of particular electrons among many others, for instance, the entanglement of quantum bits in a bath. 

\section{Conclusion}
We have investigated the real-time dynamics of inelastic electron scattering using the exact solution, TDDFT, and alternatives. The exact treatment of the dynamics reveals details in 1D e-H scattering, such as time-evolving occupation number and peaks of the energy transferred as a function of incident energy. Compared with the exact solution, ALDA substantially underestimates the energy transferred in incident energy lower than 150 eV due to the lack of strong interaction particularly in a spin-singlet state. In addition, the SIC is in a better agreement with the exact solution for ionization than the ALDA, whereas it does not correct the energy transferred at all. Therefore, it is necessary to develop an advanced XC functional capturing kinetic correlation in order to simulate accurately the inelastic scattering dynamics using TDDFT. 

The standard TDDFT in the ADLA fails to reproduce low-energy electron scattering. Nevertheless, considering the rapidly growing importance of scattering research in the physical and chemical sciences, we should make further efforts to overcome this limitation. It has been demonstrated that the two alternatives\textemdash perturbation theory and the classical limit of the incident electron\textemdash are incapable of reproducing the low-energy inelastic scattering. As a simple solution, we demonstrate a practical use of the relationship between the exact and ALDA results to correct quantitatively the ALDA result. Finally, we propose a hybrid TDDFT-TDSE method that could achieve proper electron correlation and a low computational cost simultaneously. Our work elucidates the microscopic processes in the inelastic electron-target scattering dynamics and paves the way to the study of the low-energy scattering phenomenon of larger systems with heavier atoms and more electrons in the ab initio calculation.

\begin{acknowledgments}
This work was supported by ASCENT, one of six centers in JUMP, a Semiconductor Research Corporation (SRC) program sponsored by DARPA (2018-JU-2776). This work was also supported by National Research Foundation of Korea by Creative Materials Discovery Program (2015M3D1A1068062). 
\end{acknowledgments}

%

\end{document}